\documentclass[mathpazo]{cicp}
\usepackage{algorithm}
\usepackage{algorithmic}

\begin{document}
 \title[short title]{Agent-Based Model of Chats with Bots}
\title{Can Human-Like Bots Control Collective Mood: Agent-Based Simulations of Online Chats 
}


 \author[Tadi\'c B., \v Suvakov M.]{Bosiljka Tadi\'c\affil{1}\comma\corrauth,
       and Milovan \v Suvakov\affil{1,2}}
 \address{\affilnum{1}\ Department of Theoretical Physics, Jo\v zef Stefan Institute, Box 3000, SI-1001 Ljubljana, Slovenia.\\
           \affilnum{2}\ Institute of Physics, University of Belgrade,
            Pregrevica 117, Belgrade, Serbia}
 \emails{{\tt bosiljka.tadic@ijs.si} (B.~Tadi\'c), {\tt suvakov@gemail.com} (M.~\v{S}uvakov)}


\begin{abstract} 
Using agent-based modeling approach, in this paper, we study self-organized dynamics of interacting agents in the presence of chat Bots. Different Bots with tunable ``human-like'' attributes, which exchange emotional messages with agents, are considered, and collective emotional behavior of agents is quantitatively analysed.
In particular, using detrended fractal analysis we determine persistent fluctuations   and temporal correlations in time series of agent's activity and statistics of avalanches carrying emotional messages of agents when Bots favoring positive/negative affects are active.
We determine the impact of Bots and identify parameters that can modulate it.
Our analysis suggests that, by these measures,  the emotional Bots induce collective emotion among interacting agents  by suitably altering the fractal characteristics of the underlying stochastic process.Positive-emotion Bots are slightly more effective than the negative ones. Moreover, the Bots which are periodically  alternating between positive and negative emotion, can enhance fluctuations in the system leading to the avalanches of agent's messages that are  reminiscent of self-organized critical states.
 \end{abstract}

\pac{}

\maketitle

\section{Introduction}
\label{sec1}
Recently, analysis of the empirical data from  Internet-Relayed-Chats (IRC) revealed that the self-organized dynamics of emotional messages exchanged between users may lead to social networking and recognizable collective behaviors.
In particular, the empirical data from \texttt{Ubuntu} chat channel \footnote{http://www.ubuntu.com} have been analysed as a complex dynamical system \cite{we-Chats1s, we-Chats-conference}. It has been recognized that long-term associations among users occur as a specific kind of online social network, in which both the type of messages exchanged in user-to-user communications and their emotional contents play a role \cite{we-Chats1s}.
Apart from human users,  different types of Web Bots  are often available on the channel, for example, to provide information (in terms of  ready  messages) upon user's request. Technically, chat Bots with different features can be implemented \footnote{http://www.chatbot.com}. In the current technology developments, it is tempting to use the chat Bots as mood  modifiers in online chat systems \cite{skowron-paltoglou2011afect,skowron2011}, although their  potentials to promote collective effects are currently not well understood.
From the practical point of view, recent developments in the quantitative investigation of emotions \cite{russell1980,calvo-review2010} and methods of emotion detection from a written text  \cite{Thelwall2010,paltoglou2010}, the emotion of the added Bot in a chat channel can be quantitatively defined. Practical implementation, i.e., creating the body of the messages which involve a specified emotion, is also technically feasible \cite{skowron2011,skowron2011b}. On the other hand, mechanisms that Bots can utilize to achieve their goals, as well as quantitative measures of Bot's impact on users,  depending on the properties of the Bot and the actual parameters of the system, make more difficult part of the problem.
Owing to the self-organized dynamics and appearance  of network of connections, we assume  that  particular actions of a Bot may propagate on that network and affect the course of these nonlinear processes, thus possibly altering the global state of the system. However, complete understanding of the mechanisms calls for appropriate  theoretical modeling and analysis.

In this work, we use agent-based modeling to study self-organization in the dynamics of chat  that Bots may utilize to shape the collective mood of users.  We propose  a model in which the emotional agents \cite{we_ABMblogs12,we-MySpaceABM}, representing users,  are interacting among each other and with a Bot of specified characteristics.
Bots that we introduce have several features, which make them qualitatively similar with other agents. Specifically,  Bots are inclined to direct communications, they can be picked-up by agents for conversation, and they have an emotional state which corresponds to a commonly recognized  emotion, according to Russell's model of affect \cite{russell1980}. Moreover, following the universal picture of human dynamics on the Web with characteristic delay of action (interactivity time) \cite{vazquez2006,mitrovic2010a,BT_ABMbook}, our Bots also assume a nontrivial interactivity time $\delta t_B$ for their actions.  Note that, apart from these  ``human-like'' attributes, the Bots follow some action rules which differ form ordinary agents (see details in Sec.\ \ref{sec-implementation}).

The dynamics of agent's chats consists of exchanging emotional messages on a network, which itself evolves by these chats.
By performing numerical simulations of the model for a set of control parameters which are inferred from an empirical chat system, we compute  several quantitative measures of agent's collective behaviors. Specifically,  we determine persistence in the time series of emotional messages by detrended fractal analysis. Further indicators of collective dynamics are determined from the power spectral density of these time series and the avalanches of emotional messages. 
Then we demonstrate how these quantitative measures are changed when a Bot of given emotional function is added to the system and identify the parameters that can adjust the Bot performance. We consider Bots with positive and negative emotional impact as well as the Bots that can alternate between specified emotional states.  We compare the effectiveness of these Bots by quantifying their impact on agents, and examine the nature of collective states that they provoke among the agents.

In Sec.\ \ref{sec-model} we describe  the model and provide details of numerical implementation and the control parameters. Results of the simulations in the absence of emotional Bots are given in Sec.\ \ref{sec-soc}, where the quantitative measures of self-organized dynamics are computed. In Sec.\ \ref{sec-roboteffects} the system is simulated  in the presence of a Bot with predefined positive/negative emotion valence as well as the Bots that alternate between two emotional states. Section\ \ref{sec-conclusion} contains a brief summary of the results and analysis.

\section{Model and Methods\label{sec-model}}
\hspace*{0.35cm}
We introduce an agent-based model in which agents stand for users with their attributes, which are important for the dynamics. One of the agents is identified as a Bot, which may have different properties, depending on its actual task.
We perform numerical simulations of the model maintaining  each particular message with its emotional contents, source and recipient agent, and time of the creation.  Then we analyse the simulated sequence of messages to extract the quantitative measures of agent's collective behaviors---temporal correlations and fractal properties of the time series and clustering of events (avalanches) with emotional messages. In this section, we describe the mathematical structure of the model and its numerical implementation.
We also identify the control parameters of the dynamics and determine several parameters from the empirical dataset of IRC \texttt{Ubuntu} channel.

In  analogy to real chat channel,  \texttt{Ubuntu} (data studied in Refs.\ \cite{we-Chats1s,we-Chats-conference}),  the  agents in our model interact via posting \textit{messages} on the channel. Occasionally, with a probability $g$, the posted message is directed to another agent known by its unique ID.
Thus, the agent-to-agent interactions lead to an evolving network of agents, which are interlinked by the exchanged messages.
In contrast to the bipartite network in the dynamics on Blogs \cite{mitrovic2010b}, in the IRC chats the agent-to-agent communications results in a \textit{directed monopartite network with weighted links}. On this network, each agent has specific neighborhood, on which current connections are reflecting its activity  as well as the activity of the connected neighbour agents in the preceding period.
The network evolves by the addition of new agents and/or new links among existing agents. Network evolution as well as the activity of neighbouring agents may affect the agent's emotional state, i.e., its emotional \textit{arousal}, $a_i(t)$, and \textit{valence}, $v_i(t)$. The elevated arousal may trigger agent's action---posting a new message that carries its current emotion. The precise dynamical rules  are motivated by real chat channels and are resulting in the appropriate mathematical structure of the model, to be described below.

Messages received within a given time window $T_0$ by the agent $i$ from its neighbours on the (evolving) network give rise of the influence fields $h^{a}_{i}(t)$ and $h^{v}_{i}(t)$, which can modify agent's arousal and valence, respectively (see equations below). Thus, these fields are individual properties of each agent and depend on time and actual events in the agent's neighborhood on the network.
Besides, all recent messages posted on the channel contribute to  the common fields $h^{a}_{mf}(t)$ and $h^{v}_{mf}(t)$, which  may affect  all agents.

For the agent-based modeling of chats in the presence of Web Bots, we consider the following attributes  of the agent (minimum requirements for modeling Web users are discussed in Ref.\ \cite{BT_ABMbook}):

\begin{equation}
A[id, type, N_{c}^i, g_i;  a_{i}(t), v_{i}(t), m(t), {\cal{L}}_{in}(t), {\cal{L}}_{out}(t);  \Delta t, status].\\
\label{eq-agent-bot}
\end{equation}
By adding an agent to the process, we specify its individual $id$ and $type$ (\texttt{Agent} or \texttt{Bot}). In addition, to imitate actual heterogeneity of users, i.e., due to their psychology profiles  which impact on their activity on the channel, we provide agent's capacity to create a number of messages $N_c^i$ during the simulation time as a random number taken from the empirical distribution $P(N_c)$.  We also specify the probability $g_i \in P(g)$ for the agent's preference towards identified recipient of its message, compared to writing the message on the channel, where it can be seen by all active agents.
By choosing these two distributions from the same empirical dataset, we \textit{ implicitly } fix the profile of each agent. Other approaches  are considered in the literature, e.g.,  introducing fuzzy agents with dynamical personality \cite{fuzzy_agents}.

The remaining attributes of the agent in (\ref{eq-agent-bot}) are dynamically changing with simulation time. In our model, the dynamical variables of agent's emotional state, arousal  $a_{i}(t)$ and valence $v_{i}(t)$, fluctuate under the influence of other agents and Bot.
During the chat process, each agent developes over time personal connections on the network. These are  represented by the lists of incoming and outgoing links ${\cal{L}}_{in}, {\cal{L}}_{out}$, along which the agent sent and received messages until time $t$.  Updated number of  messages posted by the agent $m(t)\leq N_c$,  as well as its lists of links and the activity \textit{status} are dynamically changing in the process of chats. The delay times $\Delta t$ is derived from the appropriate distribution whenever it is  required by the dynamic rules (see Algorithm\ \ref{alg:ABMchatswithrobotbot}). 
As it will be evident in the implementation of the model, the distinction between a regular agent and  Bot is manifested in the dynamics of their emotional states, rules of their actions, and the parameters that control their actions.

\subsection{Mathematical structure of the model\label{sec-abmstructure}}

Values of the dynamically varying fields $h^{a}_{i}(t),h^{v}_{i}(t) $, and $h^{a}_{mf}(t), h^{v}_{mf}(t))$ are determined considering the emotional arousal and valence in \textit{recently posted} messages, specifically the messages posted in the time window  $(t,t-T_0)$. 
The parameter $T_0$ is decided according to the pace of events on the channel, for instance in the \texttt{Ubuntu} channel $T_0=2$ minutes provides that in the average 10 messages are in this window.  Denoting the message $m$ creation time by $t_m$, the arousal and the valence affecting fields $(h^{a}_{i}(t)$ and $(h^{v}_{i}(t)$ of agent $i$ are determined from all recent messages directed to $i$, as follows:
\begin{equation}
h^{a}_{i}(t)=\frac{\sum_{j\in
    {\cal{L}}_{in,i}}a^{m}_{j}(\theta(t_{m}-(t-1))-\theta(t_{m}-(t-T_{0})))}{\sum_{j\in   {\cal{L}}_{in,i}}(\theta(t_{m}-(t-1))-\theta(t_{m}-(t-T_{0})))} \ , 
\label{arousal_field}
\end{equation}
where $a^{m}_{j}$ is the arousal of the message arriving along the link $j$, and
\begin{equation}
  h^{v}_{i}(t)=\frac{1-0.4r_{i}(t)}{1.4}\frac{N^{p}_{i}(t)}{N^{emo}_{i}(t)}-\frac{1+0.4r_{i}(t)}{1.4}\frac{N^{n}_{i}(t)}{N^{emo}_{i}(t)} \ .
\ , \label{valence_field}
\end{equation}
Here, the summation is over the messages from the list  ${\cal{L}}_{in,i}$ of agent's $i$ incoming links. $N^{p}_{i}$ and $N^{n}_{i}$ are the number of the messages within the considered time window which are directed to $i$  and  convey positive and negative emotion valence, respectively. $N^{emo}_{i}(t)==N^{p}_{i}(t)+N^{n}_{i}(t)$.
The corresponding components of the common fields  $h^{a}_{mf}$ and $h^{v}_{mf}$ are computed in a similar way, but considering the list of all recent messages on the channel, $\cal{S}$, including those that are not addressed to any particular agent:
\begin{equation}
h^{a}_{mf}(t)=\frac{\sum_{j\in
    \cal{S}}a^{m}_{j}(\theta(t_{m}-(t-1))-\theta(t_{m}-(t-T_{0})))}{\sum_{j\in
    S}(\theta(t_{m}-(t-1))-\theta(t_{m}-(t-T_{0})))}
\ . \label{arousal_mfield}
\end{equation}
 and 
\begin{equation}
  h^{v}_{mf}(t)=\frac{1-0.4r_{i}(t)}{1.4}\frac{N^{p}(t)}{N^{emo}(t)}-\frac{1+0.4r_{i}(t)}{1.4}\frac{N^{n}(t)}{N^{emo}(t)}
\ . \label{valence_mfield}
\end{equation}
Here, $N^{p}(t)$  and $N^{n}(t)$ is the number of positive and negative messages appearing on the channel within the current time window while $N^{emo}(t) =N^{p}(t) +N^{n}(t)$ and $r_{i}(t)=sig(v_{i}(t))$. 

Similar as in the model of blogging dynamics \cite{mitrovic2011,we_ABMblogs12} and dialogs in online social networks \cite{we-MySpaceABM}, we assume that the individual emotional state of each
agent on the chat network can be described by two nonlinear maps for the variables
$a_{i}(t) \in [0,1]$ and $v_{i}(t)\in [-1,1]$, which are appropriately coupled with the above introduced environmental fields. The rationale is that the arousal is an intensive variable, which can cause agent's action. Whereas, the positive and the negative valence have different dynamics. Consequently,  an attractive fixed point in the negative as well as in the positive valence region is expected. In the absence of psychology based mathematical model of emotion dynamics (see also \cite{schweitzer2010,BT_ABMbook}), the maps which fulfill these formal requirements include higher-order polynomial nonlinearities, specifically:
\begin{equation}
a_{i}(t+1)=
  (1-\gamma_{a})a_i(t)+ \delta(\Delta t_i)\frac{h^{a}_{i}(t)+qh^{a}_{mf}(t)}{1+q}(1+d_{2}(a_{i}(t)-a_{i}(t)^{2}))(1-a_{i}(t)),
  \label{arousal_map}
\end{equation}
\begin{equation}
v_{i}(t+1)=
  (1-\gamma_{v})v_i(t) + \delta(\Delta t_i)\frac{h^{v}_{i}(t)+qh^{v}_{mf}(t)}{1+q}(1+c_{2}(v_{i}(t)-v_{i}(t)^{3}))(1-|v_{i}(t)|).
   \label{valence_map}
\end{equation}
Here, the index $i=1,2,\ldots,N_{U}(t)$ indicates agent identity and $t$ is the time binn, which in the simulations corresponds to one minute of real time as explained in \ref{sec-implementation}. The relaxation with rate $\gamma_a=\gamma_v$ is executed systematically. Whereas, the nonlinear terms are added according to the rules, when the agent's interactivity time $\Delta t$ expires.
Note that the variables in Eqs.\ (\ref{arousal_map}) and (\ref{valence_map}) referring to the same agent are  coupled indirectly through the feedback fields and the fact that \textit{ high arousal implies potential activity}, then updated valence and arousal are transmitted in the same message.

The fields in Eqs.\ (\ref{arousal_field}-\ref{valence_mfield}) are properly normalized at each time step and the dynamical variables are kept in the range $a_i\in[0,1]$ and $v_i\in[-1,+1]$  by the maps  (\ref{arousal_map}-\ref{valence_map}). 

The sign-dependent dynamics of the valence fields in  Eqs.\ (\ref{valence_field}) and (\ref{valence_mfield})  is motivated \cite{mitrovic2011} by the nature of the valence-map in Eq.\ (\ref{valence_map}) with positive and negative fixed points and a plausible assumption that negative fields in the environment would influence more severely the agent with positive emotion than the negative one, and vice versa. The factor 0.4 in Eqs.\ (\ref{valence_field}), (\ref{valence_mfield}) is chosen to avoid unphysical fast-switching (factor 1 or 0) and contraction of phase space in the symmetric case (0.5).

\subsection{Numerical implementation of the agent-based model of chats\label{sec-implementation}}
Apart from the \texttt{Agent} and \texttt{Bot} objects defined by (\ref{eq-agent-bot}), in the simulations we keep track of each \texttt{Message} as an object $M[t_m,i,j,a_m,v_m]$ with the following attributes: creation time $t_m$, identity of the source $i$ and the target $j$ agent, and the emotional contents of the message $a_m$ and $v_m$.

Starting with an empty system, we first add a Bot. Then at each time step $t$ a number  $p(t)$ of new agents are added and animated.
The time series $p(t)$ is inferred from the empirical data of \texttt{Ubuntu} chats as the number of new arrived users (with respect to the beginning of the dataset).  Driving the agent's dynamics by the empirical  time series has the double advantage \cite{BT_ABMbook}. First, the time resolution, which is one minute in this case, sets the time scale of the simulation step, thus  providing a possibility to compare the simulated data with the empirical ones. Moreover, the empirical time series of new arrivals $p(t)$ contains circadian cycles, characteristic to human dynamics. The existence of circadian cycles is a key element of human dynamics \cite{malmgren2009} that need to be taken into account in modeling Web users \cite{BT_ABMbook}.

By the first appearance on the channel, each agent gets its fixed profile: its attitude to direct communications $g_i\in P(g)$ and the total number of messages $N_c\in P(N_c)$ that the agent can create during the simulation time $n_t$.
The distribution $P(N_c)$, $P(g)$ as well as the delay time distributions are inferred from the same empirical dataset. (In this way,  we capture potential hidden correlations between different parameters.)
On the other hand, the agent's delay time to an action, $\Delta t$, is dynamically driven from the distribution $P(\Delta t)$ after each completed action and  is systematically updated (decreased) until the condition for action $\Delta t=0$ is satisfied.

\begin{figure}[!htb]
\centering
\resizebox{24pc}{!}{\includegraphics{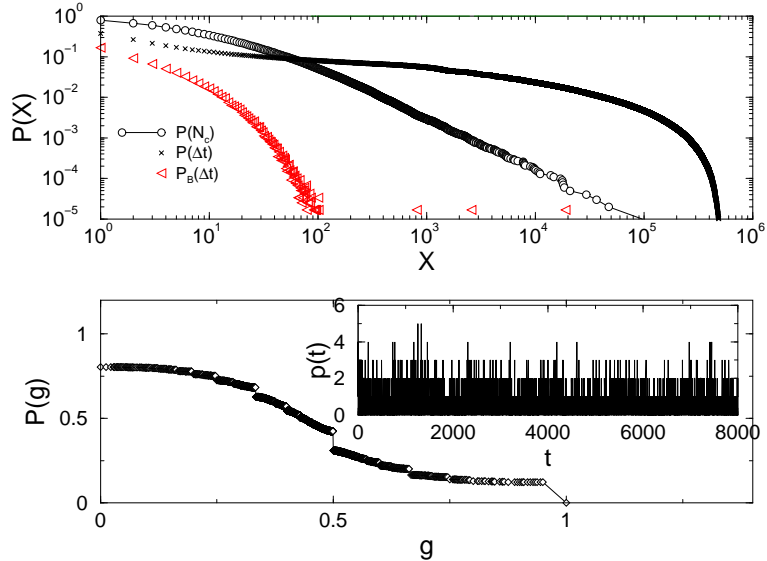}}\\
\caption{
Parameters used in the simulations are inferred from \texttt{Ubuntu} channel data: (Top) Cumulative distributions of the number of messages per user, $P(N_c)$ and user delay times $P(\Delta t)$, and distribution of Bot delay times $P_B(\Delta t)$.  (Bottom) Distribution of the probability $g$ of user-to-user message. Inset: initial part of the new arrivals time series $p(t)$ with time resolution in minutes. Total length of the time series is one year.}
\label{fig-parameters}       
\end{figure}

Designing  Bots with ``human-like'' characteristics, we assume that the Bot's interactivity time is finite. It is given by a specified distribution $P_B(\Delta t)$ with a short  average delay (see Fig.\ \ref{fig-parameters}).
We also specify $g$ value to the Bot. In this work,  we are interested in quantifying Bot's impact. For this reason, we set $g_{Bot}=0$, i.e. Bot is posting all messages to the channel. Consequently, the parameter $q$, the fraction of common-field contribution to the arousal and valence of the agents, directly moderates the importance of Bot's messages. Other interesting possibilities are, for example, $g=1$ (Bot always writes directly to an agent), or $g\in P_B(g)$, where $P_B(g)$ can be a specified mathematical or empirical distribution.

In the simulations,  we distinguish agent's status (as active or passive) according to their current participation in the dynamics. The agents arrive with a randomly chosen emotional states and their arousal and valence  between the actions are constantly decaying with the rates $\gamma_a,\gamma_v$.  The list of \textit{active agents} contains the agents who were acting or are addressed in the current time window; thus, they are likely recipients of messages from other currently active agents.
The Bot  and newly arrived  agents are automatically placed on the active list.
The interactivity time of each agent in the system is systematically monitored. The agents whose interactivity time since previous action just expires, $\Delta t_i=0$ are updated.
By computing the agent's individual fields $h^{a}_{i}(t)$, $h^{v}_{i}(t)$ and the corresponding common fields $h^{a}_{mf}(t)$, $h^{v}_{mf}(t)$  at the current time $t$, the emotional  state (valance and arousal) for each agent is updated and the  agent is then  placed to the active agents list with probability proportional to its current arousal  $a_i(t)$.
Each active agent creates a new message which, with probability $g_i$,  is directed to another agent from the active agents list; otherwise it is posted on the channel. The agents from the active list obtain new delay time $\delta t_i\in P(\Delta t)$.
The time step is completed with  decreas of their current delay times by one for all agents, and creating new active agent list.
The program flow, implemented in $C++$ code, is given in the Algorithm \ref{alg:ABMchatswithrobotbot}.

As mentioned above, the model rules are motivated by analysis of the real chat system \cite{we-Chats1s,we-Chats-conference}. Consequently, several control parameters, which are used in the simulations,  are identified and computed from the empirical data (see Table\ \ref{tab-parameters} and Fig.\ \ref{fig-parameters}).
In addition to  these empirically determined quantities, few other parameters of the model can not be estimated from the available empirical data, in particular,   the parameters of the nonlinear maps (\ref{arousal_map})-(\ref{valence_map}), relaxation rates and  the fraction  $q$ of the common field contribution. In principle, the parameters of arousal and valence dynamics can  be measured in targeted psychology experiments; however, they are not currently available in the literature. In the simulations, we fixed these parameters in the nonlinear maps such that the entire phase space can be covered when the influence fields vary within their limits. We perform simulations for different values of the parameter $q$.
List of all parameters and their values used in the simulations is given in table\ \ref{tab-parameters}.
\begin{table}[h]
 \caption{Three groups of control parameters characterize the nonlinear maps, attributes of the agents and Bots, and driving the system. The specified distributions and  numerical values are used in the simulations.}
\label{tab-parameters}
\centering
\begin{tabular}{|ll|ll|ll|}
\hline
 &Maps&                             & Agents \& Bot & &Driving \\
\hline
rate& $\gamma_a=\gamma_v=0.1$& delay time& $P(\Delta t), P_B(\Delta t)$ & agents arrival& $p(t)$\\
polynomial&  $c_2=0.5$& capacity& $P(N_c)$& mean-field& $q=0.4$\\
nonlinearity    & $d_2=0.5$& networking& $P(g)$& -fraction& $q=0.1$\\
\hline
\end{tabular}
\end{table}
As described above, the dynamic rules and parameters of the model differentiate between ``human-like'' Bot and ordinary agent by several details. Specifically, Bot's delay times are given by different distribution in which, generally, shorter delays are more probable than in the case of agents, cf. Fig.\ \ref{fig-parameters}; Its emotional state variables are not subject of the dynamic equations (\ref{arousal_map})-(\ref{valence_map}), but are either fixed values or given by another rule, depending on the task that Bot performs. Furthermore, like in the real system, Bot  is constantly presented on the channel and its capacity (number of posted messages) is not limited.
\begin{algorithm}
\caption{Program Flow: Chats of emotional agents and Bot.}
\label{alg:ABMchatswithrobotbot}
\begin{algorithmic}[1]
\STATE \textbf{INPUT:} Parameter $n_{t}$, $T_{0}$, $q$,  $a^{bot}$,$v^{bot}$; Distributions
$P_B(\Delta t)$, $P(\Delta t)$, $P(N_{c})$, $P(g)$; Time series
$\{p(t)\}$; Start list of \textit{active agents};   
\STATE Add Bot, chose $g_B$ and set $\Delta t_B=0$; 
change status active = true; 
\FORALL{$1 \leq t \leq n_{t}$}
\FORALL{$1 \leq i \leq p(t)$}
\STATE Add agent $i$; chose  $a_{i}\in[0,1]$ and $v_{i}\in[-1,1]$; chose
$g_i\in P(g)$ and $N_{c}^i\in P(N_{c})$; set $\Delta t=0$;
$N_{a}++$; add  $i$ to a list of \textit{active agents};
\ENDFOR
\FORALL{$i \leq N_{a}$} 

\STATE Calculate fields $h^{v}_{i}(t)$, $h^{a}_{i}(t)$ and $h^{a}_{mf}(t)$, $h^{v}_{mf}(t)$;
\IF{ agents delay time $\Delta t = 0$}
\STATE Update agent states $v_{i}(t)$ and $a_{i}(t)$; 
\STATE with probability $\propto a_{i}(t)$ change status active = true;
\ENDIF
\ENDFOR

\FORALL{ active agent $i$}

\IF{$i$ is Bot}
\STATE create a message $k$; transfer $v^{bot}->v^{m}_{k}$ and $a^{bot}(t)->a^{m}_{k}$;  
\ELSE
\STATE create a message $k$; transfer $v_{i}(t)->v^{m}_{k}$ and $a_{i}(t)->a^{m}_{k}$;  update $m_i(t)++$;
\ENDIF
\IF{$\xi<g$}  
\STATE pick an agent $j\in$ \textit{active agents} list;  message $k$ is referring to agent $j$; 
\ELSE
\STATE write message to the channel;
\ENDIF
\ENDFOR
\STATE clear and  update \textit{active agents} list (referring preceding $T_0$ steps); 
\FORALL{$1 \leq i \leq N_{a}$}
\STATE change status active = false;
\STATE $\Delta t--$
\IF{$\Delta t<0$ OR $i\in$ \textit{active agents} list}
\STATE chose new $\Delta t$ from $P(\Delta t)$ for agent and from
$P_B(\Delta t)$ for a Bot
\ENDIF
\ENDFOR
\ENDFOR
\STATE {\bf END}
\end{algorithmic}
\end{algorithm}

\section{Results and Discussion\label{sec-results}}
\begin{figure}[!h]
\centering
\begin{tabular}{cc}
\resizebox{32.8pc}{!}{\includegraphics{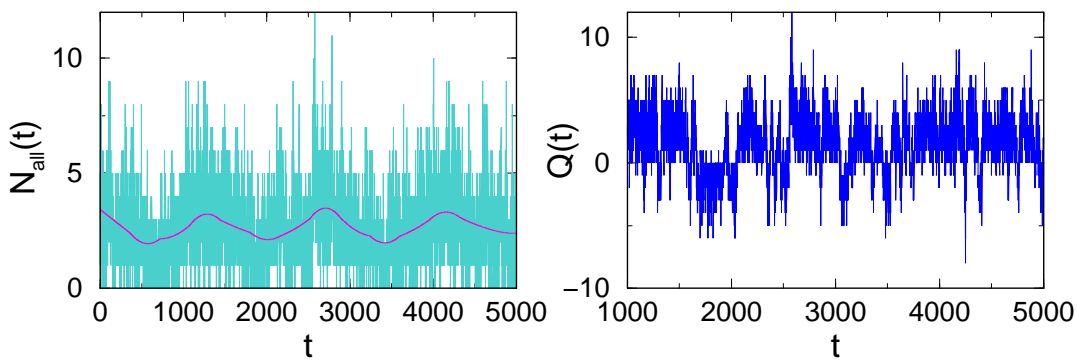}}&
\end{tabular}
\caption{Simulated time series of messages in the absence of Bot: (a) All messages $N_{all}(t)$  and trend signal with daily cycles and (b) charge of emotional messages $Q(t)= N^+(t)-N^-(t)$ are plotted against time  $t$.}
\label{fig-ts0}       
\end{figure}

\hspace*{0.35cm}
Our simulated data are stored as a sequence of events, which describe in detail each message created by agents and Bot: creation time, sender agent, recipient agent or channel, and arousal and valence that the message carries.
To analyse these data, we first construct several time series and agent's associations as a  network. In particular, by respecting specific attributes of these messages, we consider:
\begin{itemize}
\item $N^\pm(t)$: Time series of the number of messages at time step $t$ carrying positive/negative emotion valence;
\item $N_{all}$: Time series of the number of all messages at time $t$, irrespective their emotion;
\item $Q(t)$: Time series of the charge of emotional messages, $Q=N^+(t)-N^-(t)$; 
\item $C_{ij}(T_W)^\pm$: Network connections among agents that emerge within a specified time window $T_W$; Links carrying positive emotion messages can be  differentiated from the negative messages links; 
\end{itemize}
For the illustration,  in Fig.\ \ref{fig-ts0}a we show a part of the time series $N_{all}(t)$ for the first 5000 steps, corresponding roughly to three and half days. Notably, the time series exhibits daily cycles of agents activity, which is induced by the corresponding cycles in the driving signal $p(t)$ (see inset to Fig.\ \ref{fig-parameters}).
The corresponding time series of the number of positive/negative messages $N^\pm(t)$ are the respective subsets of $N_{all}(t)$. Consequently, $N^\pm(t)$ also exhibit daily cycles; however, their difference does not have such periodicity.
The excess of the messages of one or the other polarity is defined as ``charge'' of emotional messages \cite{warsaw2011,mitrovic2011}. Simulated in the absence of Bots, the charge of  agent's messages appear to fluctuate around zero as shown in Fig.\ \ref{fig-ts0}b. In contrast to the original time series,  $N^+(t)$ and $N^-(t)$, the time series of charge does not exhibit  any prominent cycle. These conclusions are  supported by the detrended time series analysis in Sec.\ \ref{sec-soc} and \ref{sec-roboteffects}.

\begin{figure}[!h]
\centering
\begin{tabular}{cc}\resizebox{24pc}{!}{\includegraphics{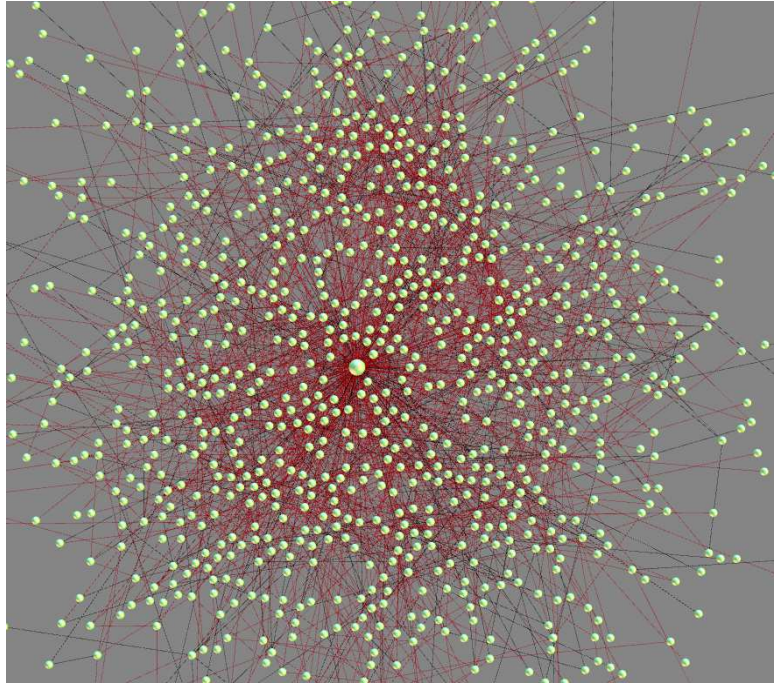}}\\
\end{tabular}
\caption{Network that emerges after 2000 steps in direct exchange of emotional messages between agents in the presence of  \texttt{joyBot}. Links with mostly positive (red) and negative (black) emotion messages are indicated. }
\label{fig-hBotnetwork}       
\end{figure}

We build the emergent network of chats from the simulated list of messages, by considering directed agent-to-agent, agent-to-Bot and  Bot-to-agent  communications and disregarding the messages that are posted on the channel without specified recipient. For the purpose of this work, the  network that emerges after the first 2000 steps in the presence of a positive-emotion Bot is shown in Fig.\ \ref{fig-hBotnetwork}. It contains approximately 1000 agents who were involved in personal communications.  The red color of links indicates positive messages exchanged along them while the black color stands for negative emotion messages. As one can see, a majority of  links among the agents even away from the Bot is positive, suggesting that the emotion impact of the Bot spreads over the network.
The chat network belongs to the class of hierarchically organized structures, similar to the ones inferred from the empirical chat data \cite{we-Chats1s,we-Chats-conference}. A detailed analysis of such networks, their multi-relational structure, dependence on the parameters of the model and Bot's activity, as well as comparisons with real chat systems are studied in a separate paper \cite{we-Chats-chapter}.

\subsection{Self-Organized Behavior of Emotional Agents\label{sec-soc}}

\begin{figure}[!htb]
\centering
\resizebox{34.8pc}{!}{\includegraphics{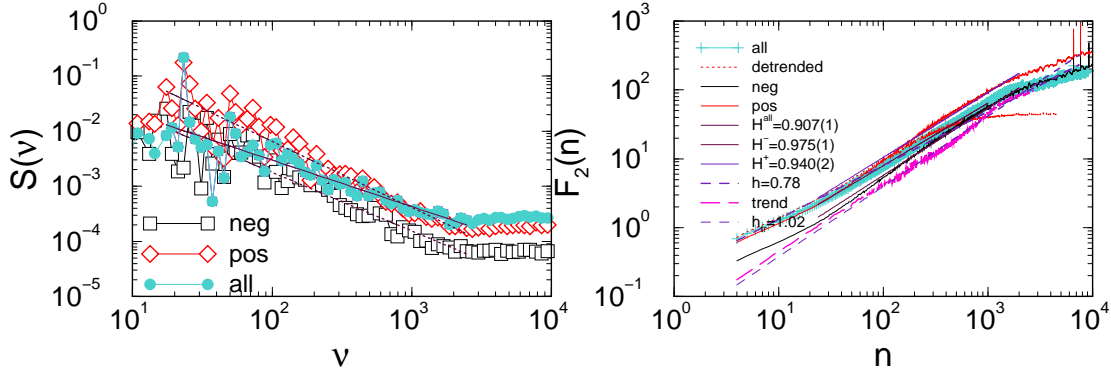}}\\
\caption{Simulated time series of messages with positive $N^+(t)$ and negative $N^-(t)$ valence and all messages $N_{all}(t)$ in the absence of Bots: (a) power spetrum and (b) fluctuations of these time series and of the trend. }
\label{fig-tsPS-H}       
\end{figure}

In analogy with the empirical  data of \texttt{Ubuntu} chats \cite{we-Chats1s,we-Chats-conference}, the time series of chats simulated by our agent-based model exhibit characteristics of fractal stochastic point processes. This implies that different quantitative measures follow scaling behavior within extended time windows \cite{fractalSPP}. In particular, the power spectral densities, shown in Fig.\ \ref{fig-tsPS-H}a for different simulated time series, obey  power-law decay $S(\nu)\sim \nu ^{-\phi}$ with the scaling exponents close to the flicker noise. We find $\phi ^{+}=1.06\pm 0.06$, for positive and $\phi^{-}=1.19\pm 0.06$ for negative emotion messages time series. While, for the signal containing all messages  irrespective their emotional polarity, we obtain $\phi=0.84\pm 0.06$, which is compatible with the fractal Gaussian-noise driven Poisson process \cite{fractalSPP}.

Furthermore, we analyse persistence of the fluctuations in the integrated time series at varying time interval $n$, measured by the Hurst exponent.
Note that these time series are stationary but obeying daily cycles (and potentially higher cycles), which intend to increase persistence \cite{gao_cycles}.
To remove the cycles, we adapted detrended fractal analysis with polynomial interpolation, described in detail in Ref.\ \cite{we-MySpace11}. An example of local trend signal with daily cycles of time series messages is shown in Fig.\ \ref{fig-ts0}a. For a time series $h(k), k=1,2,\cdots T$, the profile $Y(i)=\sum_{k=1}^i(h(k)-<h>)$ is divided into $N_n$ segments of length $n$ and the scaling of the fluctuations against segment length computed as follows: $F_2(n)=\left[(1/N_n)\sum_{\mu=1}^{N_n}F^2(\mu,n)\right]^{1/2} \sim n^H$.  Here, the fluctuation at $\mu$-th segment  $F^2(\mu,n)= (1/n)\sum_{i=1}^n[Y((\mu-1)n+1)-y_\mu(i)]^2$ is the standard deviation from the local trend $y_\mu(i)$. The results for several time series (indicated in the legend) are shown  in Fig.\ \ref{fig-tsPS-H}b.  Note that  the obtained Hurst exponents for all these time series are in the range $H\in(0.78,1.0)$, suggesting \textit{strongly persistent} fluctuations in the time interval shorter than one day (scaling region).

The observed  fractality in the time series of chats suggests that clustering of events over larger time scales occurs in the process. 
The self-organized nature of the dynamics is further tested by analysis of the \textit{avalanches} of temporally connected events \cite{dhar1990,jensen1998,corral2004,tadic1997,tadic1999}. For the purpose of this work, we consider the avalanches of emotional messages. To determine avalanches from the time series of events,  we apply the methodology, which is used in the analysis of Barkhausen noise signals \cite{spasojevic1996,djole2011} and avalanches of comments in the datasets from  Diggs \cite{mitrovic2011}. In a time signal, the avalanche comprises of the number of events  between two consecutive points where the signal meets the base line (here zero level). Then the distance between these two points determines  the duration of the avalanche, $T$, while the amount of messages between these two points is the avalanche size, $s$.

\begin{figure}[!htb]
\centering
\resizebox{32.8pc}{!}{\includegraphics{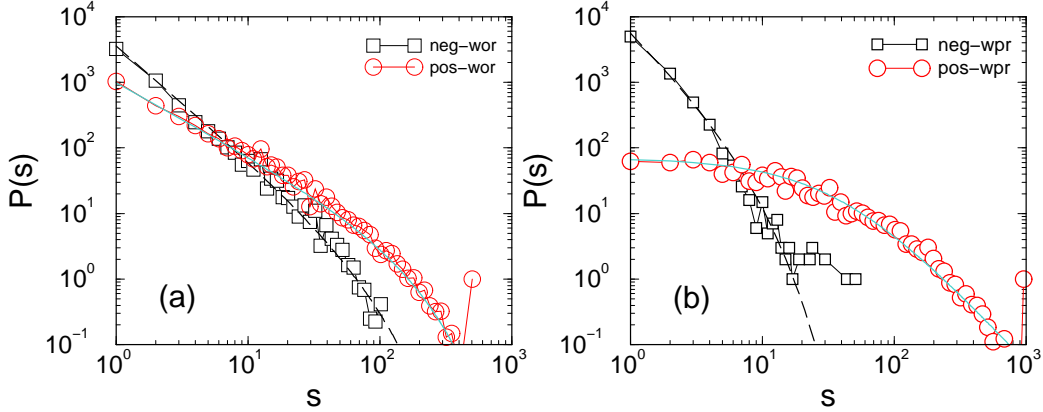}}\\
\caption{Distributions of size of avalanches with negative and positive emotion messages of agents in the absence of Bot (left panel), and  when \texttt{joyBot}  favoring positive emotions is present (right panel).
}
\label{fig-avalanches-pr}       
\end{figure}

The distributions of avalanche sizes determined from the signals of messages carrying positive and negative emotional valance are shown in Fig.\ \ref{fig-avalanches-pr}a. Noticeable, the distributions of  sizes of the emotional avalanches (and likewise the distributions of durations, not shown) obey a power-law dependence before a cut-off $s_0$,
\begin{equation}
P(s) = Bs^{-\tau_s}\exp{-(s/s_0)}  \ ,
\label{eq-size}
\end{equation}
as expected for the self-organized systems of finite size \cite{spasojevic1996,tadic1997}. We find
different exponents $\tau_s^- =1.83\pm 0.04$ and $\tau_s^+ =1.34\pm 0.03$ for  the avalanches carrying negative  and positive emotions, respectively.
The duration exponents $\tau_T^{\pm}$ are in the similar range; the avalanche shape exponents $\gamma_{ST}^\pm= \frac{\tau_T^\pm-1}{\tau_s^\pm-1}$ are found as follows: $\gamma_{ST}^+=1.47$ and $\gamma_{ST}^-=1.23$. The larger exponent (broader shape) of positive avalanches suggests  that in propagating positive emotion branching more often occurs than  when  negative emotion is transmitted between agents.
In the following section,  we examine how these self-organized mechanisms of chats are altered when a mood-modifying Bots are available.

\subsection{Collective Effects of Mood-Modifying Bots\label{sec-roboteffects}}
\begin{figure}[!htb]
\centering
\includegraphics[scale=.48]{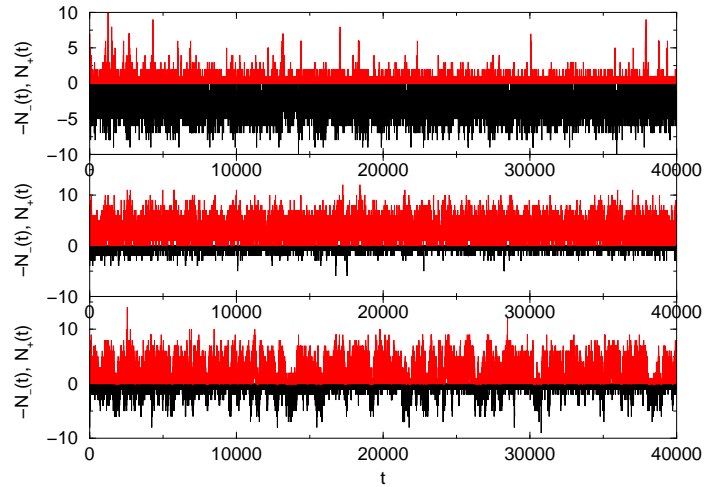}
\caption{Time series of te number of positive $N^+(t)$ and negative $N^-(t)$ messages without Bots (bottom) and in the presence of \texttt{joyBot-q04} (middle) and \texttt{misBot-q04} (top). For better vision $-N^-(t)$ is plotted.
}
\label{fig-ts-pr}      
\end{figure}

\hspace*{0.35cm}
To activate a Bot in the system, here, we first define Bot's role as a possible mood modifier. In this work, we are interested in understanding potentials of emotional Bots to promote   collective reaction among  interacting agents with dominant positive or negative emotion, and in quantifying the effectiveness of Bots. Therefore, we first consider two Bots with fixed emotion, corresponding to common emotions known as ``joy''  and ``misery'', respectively. According to  the Russell's two-dimensional circumplex model \cite{russell1980}, these are two emotions with opposite valence and similar arousal. In addition, we define  a  Bot that can dynamically alternate between these two emotions. Specifically, we simulate the system of agents in the presence of following Bots:
 \begin{itemize}
\item \texttt{joyBot-q04} is constantly active with the emotion ``joy'' $(a^*=0.5, v^*=+1)$;
\item \texttt{misBot-q04} is constantly active with the opposite emotion ``misery'' $(a^*=0.5, v^*=-1)$;
\item  \texttt{altBot-q04} and \texttt{altBot-q01} are alternating between $(a^*=0.5, v^*=-1)$ and $(a^*=0.5, v^*=+1)$ every three days (4320 simulations steps); The parameter values $q=0.4$ and $q=0.1$ indicate that the effectiveness of Bot's messages on the channel is modified.
\end{itemize}
Performing simulations of agents activity in the presence of emotional Bots, we determine the time series of positive and negative emotional messages at the level of the entire system. Beginning parts of these time series are displayed in Fig.\ \ref{fig-ts-pr}; They point out prevalence of positive emotion messages when \texttt{joyBot} is active, and similarly, excess of negative valence emotions in the case  of  \texttt{misBot}. For illustration, in the bottom panel  we display the time series in the case  without Bots. These findings are in agreement with the structure of links in the emergent network in Fig.\ \ref{fig-hBotnetwork}, where one can see that a majority of links carry positive emotion (indicated by red color) between the agents, including those which are not in a direct contact with the \texttt{joyBot-q04}.

\begin{figure}[!htb]
\centering
\includegraphics[scale=.48]{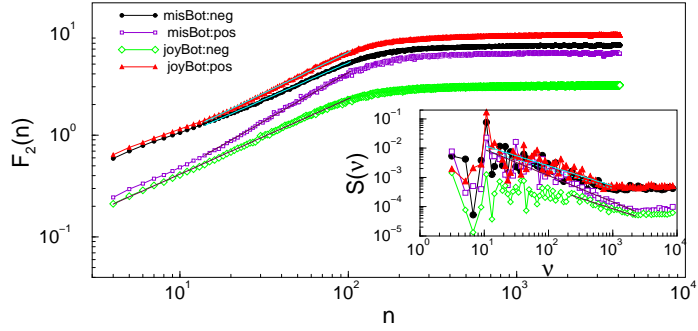}
\caption{Fluctuations of the agent's emotional messages  time series in the presence of \texttt{misBot-q04} and \texttt{joyBot-q04}.
}
\label{fig-PS-H_twoBots}      
\end{figure}
The fractal structure of these time series is also changed, compared to the emotional series in the absence of any Bots. Selecting out all messages posted by Bots, we evaluate emotional time series from the remaining data of agent's messages.  The results for the power spectral densities $S(\nu)$ vs. $\nu$ and fluctuations $F_2(n)$ vs. $n$ of agent's messages with positive and negative valence in the presence of two emotional Bots are shown in Fig.\ \ref{fig-PS-H_twoBots}.
Specifically, when the negative Bot \texttt{misBot-q04} is active, the scaling exponents for negative and positive emotion messages of agents are found $\phi^{--}=0.66(7)$, $H^{--}=0.721(3)$  and  $\phi^{-+}=0.89(6)$, $H^{-+}=0.952(5)$, respectively. (Here, the first index denotes Bot's emotional polarity while the second index stands for the polarity of agent's messages.)
Similarly, when the positive Bot \texttt{joyBot-q04} is present the fluctuations of the positive and negative emotion messages differ, leading to the exponents  $\phi^{++}=0.68(8)$, $H^{++}=0.799(3)$, and $\phi^{+-}=0.71(5)$, $H^{+-}=0.741(3)$. The corresponding ranges where the scaling holds are indicated by the straight line along each curve in Fig.\ \ref{fig-PS-H_twoBots}.  Note that a typical cycle in these time seres was  removed, resulting in the plateau for $n>100$.
The slopes of the fluctuation curves increase in the range $n\in[10,100]$, which can be related with the impact of Bots.
It is interesting to note that generally, stronger fluctuations in the emotional time series (larger Hurst exponents) are observed when neutral Bot or Bots with opposite emotional polarity are active. Whereas, weakest fluctuations are found in the emotional messages matching Bot's polarity.  We have  $H^{--} \lesssim H^{+-} < H^{0-}$, for negative emotion messages, and $H^{++}<H^{-+} \lesssim H^{0+}$, for positive emotion messages.

Further effects of emotional Bots are found in the distribution of avalanche sizes and durations. As  it can be expected, the cut-off sizes increase for the avalanches of emotional messages matching Bot's polarity. In Fig.\ \ref{fig-avalanches-pr}b, we show the results for 
the distribution of size of positive and negative emotion avalanches 
in the presence of \texttt{joyBot-q04}. The cut-off size of positive avalanches in creases,  
while the size of negative avalanches decreases, compared with the situation when Bot is absent, cf. Fig.\ \ref{fig-avalanches-pr}a. 
More importantly,  the distribution of the enhanced positive avalanches
manifests different mathematical curve. In precise,  the distribution can be properly approximated by $q$-exponential expression
\begin{equation}
P(s)=C(1-(1-q)s/s_0)^{1/1-q} \ ,
\label{eq-size-q}
\end{equation} 
when Bot is active, compared with the Levy-type distribution with cut-off in Eq.\ (\ref{eq-size}), which is characteristic for the avalanches in the absence of Bots.  In contrast, the frequency of avalanches of negative messages is drastically reduced and obeys exponential decay. Qualitatively similar effects are observed, with the interchanged signs of emotions, in the presence of \texttt{misBot-q04}.

Theoretically, the $q-$exponential distribution \cite{tsallis1996} of the type \ref{eq-size-q} is anticipated in dynamical systems with  reduced phase space \cite{ST_entropiesEPL}, e.g., accelerating random walk \cite{ST_entropiesEPL}, superdiffusion on networks \cite{tadic2004,TT04} and other.
Similar type of distributions has been derived for the avalanches in non-interacting dynamical systems exposed  to a coherent noise  \cite{Sneppen96coherentnoise}.
Although the presence of Bots with a constant emotion can be considered as a coherent input directed to a large number of agents (but not all of them), it should be stressed that the agents in our model are interacting. The interactions among agents have profound effects as above demonstrated. Specifically, the system developed temporal correlations and fractal time series, which is not the case with non-interacting agents  in Ref.\ \cite{Sneppen96coherentnoise}.
In the following, we will consider Bots with time dependent inputs, \texttt{altBot-q04} and \texttt{altBot-q01}. We will also show that they produce a distinctive impact on the avalanches, in contradiction to the constant-emotion Bots.

\begin{figure}[!htb]
\centering
\resizebox{24pc}{!}{\includegraphics{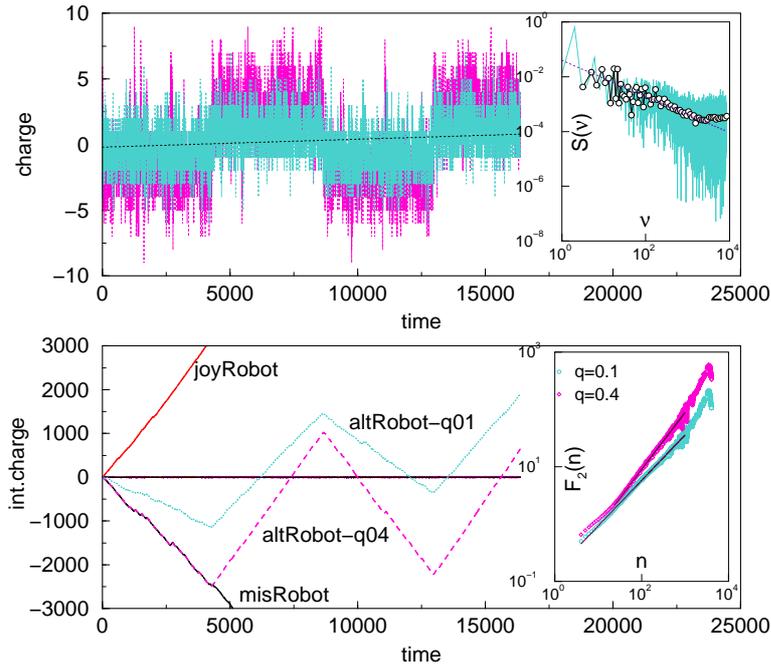}}\\
\caption{Effects of the alternating-mood Bots for two values of the parameter $q$=0.1 and 0.4: Charge of the emotional messages of agents (top) and integrated charge (bottom panel) plotted against time. Insets: Power spectral density (top) and fluctuations (bottom).
}
\label{fig-ts-altr}       
\end{figure}

\textit{Bots with time-dependent inputs} are suitable to  examine how reliable and effective is the impact of Bots on the overall emotional state of agents. Here, we analyse the simulation data for the case of periodically alternating Bots, \texttt{altRobot-q04} and \texttt{altRobot-q01}. As above, by selecting out  the messages posted by the Bot, we consider the remaining data which comprise the agent's activity when the Bot was present.  In Fig.\ \ref{fig-ts-altr}(top), we show the charge of the emotional messages by agents in the presence of alternating Bot.
The time series of emotional charge exhibits excess negative/positive values following the switching of Bot's polarity.
The two time series are for the charge $Q(t)$ of the emotional messages of all agents'  in the presence of alternating Bot when the parameter $q$ is varied as $q=0.4$, corresponding to strong Bot's influence (pink line) and $q=0.1$, decreased influence of Bot's messages (cyan).
Note that the  time series of charge induced by the alternating Bot also shows long-range correlations. For instance, by computing the power spectrum and fluctuations of the charge time-series in the case of \texttt{altRobot-q01}, we find the following scaling exponents:  $\phi^Q=0.66(8)$ and $H^Q=0.791(2)$. The results are shown in the insets to Fig.\ \ref{fig-ts-altr}.

In order to demonstrate how quickly the excess positive/negative charge is built in the agent's emotional messages, we introduce the normalized charge defined as
\begin{equation}
Q_n(t)=\frac{N_+(t)-N_-(t)}{N_+(t)+N_-(t)} \ .
\label{eq-normQ}
\end{equation}
 Then, the integrated normalized charge signal $iQ_n(t)=\sum_{t^\prime=1}^tQ_n(t^\prime)$ increases/decreases with time, with the rate which is given by the slope of the curve $iQ_n(t)$.  In the bottom panel of Fig.\ \ref{fig-ts-altr}, $iQ_n(t)$ is plotted against time $t$ for four Bots studied in this work.
As the Fig.\ \ref{fig-ts-altr} shows, when the \texttt{joyBot-q04} is present, the excess positive charge is building quite efficiently. Likewise, the excess negative charge builds when the \texttt{misBot-q04} is active. Comparison of the slopes of these two curves, +0.808 and -0.714, indicates that the positive Bot is a bit more effective than the negative one. Note that all other parameters of the system as well as the sequence of random numbers are the same.
When the Bot is alternating, starting from the negative episode, the induced total charge of agent's emotional messages also oscillates with the same period. The respective sections of the  $iQ_n(t)$ curve have the same slopes as the case with negative/positive Bots.  Whereas, the corresponding slopes are smaller, resulting in a shallow curve in the case $q=0.1$. This  indicates that the alternating Bot for $q=0.1$ is less effective, compared with the one for $q=0.4$. Owing to slight  asymmetry of the charge signal, the overall curve is slowly shifting towards positive charge.

\begin{figure}[!htb]
\centering
\resizebox{32.8pc}{!}{\includegraphics{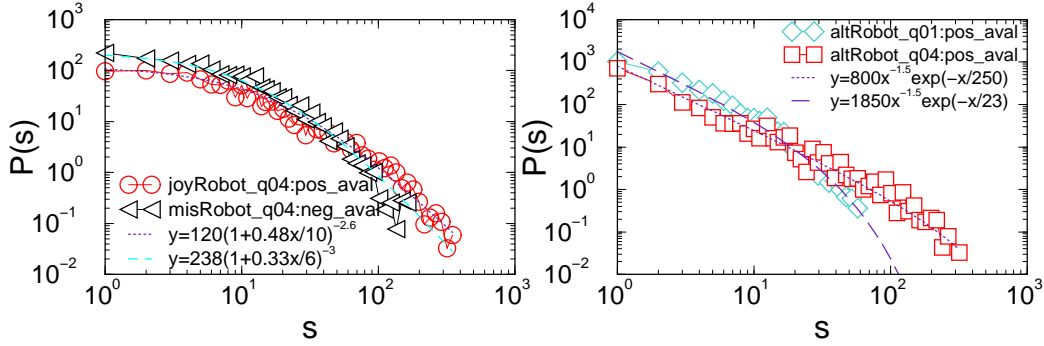}}\\
\caption{Comparison of the distributions of sizes of the emotional avalanche of  agent's messages in the presence of different emotional Bots: (a) fixed positive and negative emotion Bots with the same parameter $q=0.4$, and (b) alternating Bots with varied parameter $q=0.4$ and $q=0.1$
}
\label{fig-avalanches-compare}       
\end{figure}

Further quantitative measures of the effectiveness of  alternating Bots can be inferred  from the analysis of \textit{ persistence } of the charge fluctuations and \textit{ avalanches of emotional messages }.
In the inset to Fig.\ \ref{fig-ts-altr} (bottom panel), a larger Hurst exponent $H$ is found for the fluctuations of charge time series when the Bot's effects are stronger ($q=0.4$) as compared with the case when the Bot's messages are not as much recognized by the agents ($q=0.1$). In the latter case, we find Hurst exponent $H=0.791\pm 0.002$ in the time window $n\in[4:1000]$, while in the first $H=0.992\pm 0.004$ is found and applies in the area $n\in [20:1000]$ below the Bot-induced periodicity (n=4320).
The activity of the alternating-mood Bots tends to enhance long-range correlations in the system. Appart form the persistence of the charge fluctuations this tendency can be also seen in the altered nature of avalanches.
In Fig.\ \ref{fig-avalanches-compare}a, the distributions of  avalanche sizes induced by the Bot of the same polarity: positive emotion avalanches induced by the behavior of \texttt{joyBot-q04} and negative-emotion avalanches in the presence of  \texttt{misBot-q04} with the same $q$ parameter. Generally, larger positive avalanches can be observed in the presence of positive Bot, than negative-emotion avalanches when the negative Bot is available.

The presence of alternating Bots  enhances fluctuations in the agent's emotions, as already mentioned. Here, in Fig.\ \ref{fig-avalanches-compare}b we show that the structure of emotional avalanches is also altered, compared with the fixed-emotion Bots.
The power-law distributions with the universal exponent $\tau_s=1.5$ is found both for positive and negative emotion avalanches in the presence of the alternating Bots. When the alternating Bot is  more effective, i.e., for $q=0.4$, we find increased probability of  very large avalanches,  compared with $q=0.1$.

Apart from the parameter $q$, which measures how Bot's messages are taken into account in the common influence fields, Bot's activity can be tuned by varying its delay time $\Delta t_B$.  In the present work, the characteristic delay of Bots $<\Delta t_B> \sim 10$, cf. distribution $P_B(\Delta t)$ in Fig.\ \ref{fig-parameters}. According to the dynamic rules, the impact of Bot generally depends on the activity of the agents. Having the simulated data, we can estimate Bot's impact as follows:
\begin{equation}
B_a=\left[\frac{1}{<N_{active}>+1}\frac{t}{<\Delta t_B>}\right]q \ ,
\label{eq-Bot-rate}
\end{equation}
where the average number of active agents $<N_{active}>$ affects the  likelihood that  Bot is picked up for discussion. Conversely, increased activity of  Bot may induce a larger number of active agents. In the simulated data with two alternating Bots, we have $<N_{active}>=$ 2.478 and 1.441, corresponding to  $q=$0.4 and 0.1. Thus, respective Bot's activity rates are estimated as $B_a/t=$ 0.012 and 0.004.

\section{Conclusion\label{sec-conclusion}}
\hspace*{0.35cm}
We have introduced an agent-based model of online chats with the presence of emotional Bots. The properties of the agents and the model rules are motivated by user dynamics observed \cite{we-Chats1s,we-Chats-conference} in the analysis of empirical data in IRC \texttt{Ubuntu} channel. Several control parameters of the model are inferred by analysis of the same empirical data, and the system is driven by the empirical time series---arrival of new users (agents).

First, by shutting down the Bots, we have analysed the underlying stochastic process of emotional interactions among the agents.
Systematic analysis of the simulated time series of agent's activity in the absence of Bots reveals the self-organized nature of the dynamics. Several quantitative measures that we determine  imply the fractal Gaussian-noise driven processes. The persistent fluctuations, correlations in the time series and power-law avalanches carrying positive and negative emotion messages are determined as quantitative characteristics of collective behaviors of agents. The observed temporal correlations, as well as the emergence of the network of agents, share striking similarity with the empirical system \cite{we-Chats1s,we-Chats-conference}.

Furthermore, by animating emotional Bots with ``human-like'' characteristics, we have shown that the character of the underlying fractal process is altered, enabling Bot's impact on collective emotional behaviors of agents.
In particular, we have demonstrated that Bots with fixed emotional state (opposite emotion ``joy'' and ``misery'' are analysed), interacting with a certain number of agents, induce dominant emotional valence among the agents in the whole system. Detailed study of the persistence, correlations and clustering (avalanche) of emotional messages of agents, suggests that Bot's activity induces changes  in the quantitative measures of the process and reduction of the phase space. Based on the self-organized nature of the process, these mechanisms guarantee  reliable effects  of Bots activity. Quantitative measures of Bot's impact depend on several parameters of Bots and the interacting agents.  For the set of parameters used in our simulations, we find that, in general, processes building positive emotional states are more effective than the ones with negative collective response.

Moreover, our simulations revealed that the presence of Bots which are alternating between positive and negative mood enhances fluctuations and further changes temporal correlations in the emotional messages at the level of the whole system. When the mood-alternating Bot is extremely effective, as the case of \texttt{altBot-q04}, the fractal characteristics of the process approach  the ones of flicker noise and the avalanches resembling self-organized critical behavior.
These theoretical aspects, as well as design of Bots with different functions, are left for future work.

\section*{Acknowledgments}
We would like to thank the support from the program P1-0044 by the  Research agency of the Republic of Slovenia and from the European Community's program  
FP7-ICT-2008-3 under grant n$^o$ 231323. We would also like to thank V. Gligorijevi\'c for useful comments.



\end{document}